\documentclass[11pt]{article}
\usepackage{jheppub}
\usepackage{amsmath,amsopn,amssymb,amsfonts,amsthm,mathrsfs,bbm,caption,latexsym,stmaryrd,verbatim,color,graphicx}
\usepackage{hyperref}
\usepackage[dvipsnames]{xcolor}

\let\l\left
\let\r\right

\let\mb\mathbb
\let\mc\mathcal

\let\mg\mathfrak

\newcommand{\nn}{\nonumber\\}

\title{Localization of a supersymmetric gauge theory in the presence of a surface defect}
\author{J. Lamy-Poirier}
\affiliation{Perimeter Institute for Theoretical Physics, 
Waterloo, Ontario, Canada N2L 2Y5}
\affiliation{Department of Physics and Astronomy,
University of Waterloo, Ontario, Canada N2L 3G1}

\abstract{We use supersymmetric localization to compute the partition function of $\mc N=2$ super-Yang-Mills on $S^4$ in the presence of a gauged linear sigma model surface defect on a $S^2$ subspace. The result takes the form of a standard partition function on $S^4$, with a modified instanton partition function and an additional insertion corresponding to a shifted version of the gauged linear sigma model partition function.}

\begin{document}

\maketitle
\flushbottom


\newpage

\section{Introduction}
\label{sec:intro}

Supersymmetry, despite being so far of little use for realistic physical models, is an interesteresting framework for studying nonperturbative phenomena in quantum field theory. Indeed, many physical quantities invariant under some supersymmetry can be computed exactly using various tools. One such tool is localization \cite{Witten:1988ze}, which allows to compute some partition functions and supersymmetric quantities by deforming the action in a suitable way. In recent years, localization allowed to compute partition functions and Wilson loops of supersymmetric gauge theories on $S^4$ and its deformations \cite{Pestun:2007rz,Gomis:2011pf,Hama:2012bg}, on $S^2$ \cite{Doroud:2012xw,Benini:2012ui}, and on various other spacetimes. In two dimensions, the method has also proven successful in computing various other physical and mathematical quantities \cite{Doroud:2013pka,Gomis:2012wy,Jockers:2012dk,Hori:2013ika,Benini:2013xpa,Ashok:2013pya}.

In this paper, we explore another direction in which the localization program can be expanded, that of supersymmetric gauge theories with surface defects. Surface defects play an important role in quantum theory, and therefore are interesting objects to study \cite{Gaiotto:2009fs,Drukker:2010jp}. Our goal is to localize a $\mc N=2$ supersymmetric gauge theory on $S^4$ interacting with some extra GLSM degrees of freedom on a $S^2$ surface\footnote{It would also be possible include twisted multiplets and perform localization as in \cite{Doroud:2013pka}, but we do not explore this possibilty here}. The surface is embedded as a great sphere of $S^4$ (i.e. it contains antipodal points). The presence of the defect breaks half the supercharges, leaving a $\mc N=(2,2)$ supersymmetry group. We also focus on the partition function of that theory. In addition to being an interesting result by itself, the partition function plays an important role in the AGT duality \cite{Alday:2009aq,Alday:2009fs}, and thus its computation allows for an additional check of the conjecture. The computation also sets the ground for that of other physical quantities (compatible with localization) in the same setup. For most of the paper, we specialize the computation to the case of a pure supersymmetric gauge theory on $S^4$ interacting with chiral multiplets on the defect. The simpler computation already shows all the important features of the general case, allowing for a straightforward generalization. In particular, charging a chiral multiplet under a 2d representation of a 4d vector multiplet gives the only coupling between the bulk and the defect relevant for the computation. Superpotential interactions are also possible, but they do not directly affect the path integral.



We compute the partition using supersymmetric localization in the Coulomb branch. The result is an integral over the coulomb branches of both the 4d and 2d multiplets, of the form:
\begin{align}
Z_{2d-4d}&=\int_{4d\text{ Coulomb}}da~e^{-S^{4d}_{cl}(a)}Z^{4d}_{1-\text{loop}}(a)
\nn
&\qquad\times \int_{2d\text{ Coulomb}}d\tilde a~e^{-S^{2d}_{cl}(\tilde a)}|Z_{\text{inst}}^{2d-4d}(\tilde a,a)|^2
Z^{\text{chiral}}_{1-\text{loop}}(\tilde a,a)Z^{2d\text{ vector}}_{1-\text{loop}}(\tilde a).
\end{align}
(See eq. (\ref{finalResult}) for the exact formula.) The one-loop determinants are the same as in the partition function of the isolated 2d and 4d theories, except for the chiral multiplets which see the 4d vector multiplets as background vector fields. The instanton partition function corresponds to a modified version of Nekrasov's partition function, where the gauge theory in the $\Omega$-background now contains a $\mb R^2$ defect. It is expected to be computable using the methods of \cite{GaiottoKim}, but its derivation is left for future work.



The paper is organized as follows. In section \ref{sec:theory}, we describe the theory theory of a 2d chiral multiplet and a 4d vector multiplet and its relevant properties. The relevant computations and some technical details are left for the appendices. The localization is performed in section \ref{sec:loc}, where we find the appropriate setup for the localization and assemble the components of the partition function. Most of the components can be taken from the literature with little change, and we review the computation of the one-loop determinant for the chiral multiplet in appendix \ref{one-loop}. In section 
\ref{sec:general}, we show how the results generalize in the presence of matter on $S^4$ and vector multiplets on the defect.

\section{The theory}
\label{sec:theory}

In this section we write down the action and some basic properties of the theory of a vector multiplet on $S^4$ coupled to a chiral multiplet on a $S^2$ defect. The separate uncoupled theories on $S^2$ and $S^4$ have been described previously \cite{Doroud:2012xw,Benini:2012ui,Pestun:2007rz}. Our goal is to find the coupling between the defect and the four dimensional vector multiplet. For this purpose, we express the vector multiplet in a ``two dimensional" language, where only the unbroken symmetries of the theory are manifest. We split the tangent space near the $S^2$ subspace in a parallel and a perpendicular part, and we write the spinors as two dimensional Dirac spinors (relative to the parallel part of the tangent space). Then the restriction of the multiplet to the $S^2$ subspace can be seen as a set of field on that subspace, hence it can interact locally with the matter on the defect. In particular, the derivatives of the fields in the orthogonal direction appear as an extra set of fields. In the following, we call this process the ``restriction" of the vector multiplet from $S^4$ to $S^2$.

When restricted to $S^2$, the four dimensional vector multiplet appears as a tower of fields, which consists of the vector multiplet on the defect and its transverse derivatives. Each level appears as a two dimensional vector multiplet together with a chiral multiplet. In principle we can couple the defect to any of this fields, but renormalizability forbids any supersymmetric to the derivative fields (see appendix \ref{restriction}).

The resulting chiral multiplet is coupled to the vector multiplet in an unusual way, which takes care of the dependence in the transverse dimensions. However, the (restricted) vector multiplet itself appears exactly as a $\mc N=(2,2)$ vector multiplet, with the correct supersymmetry transformations. Therefore we can couple the 2d matter to the restricted vector multiplet as if it was a vector multiplet on $S^2$, and the result is guaranteed to be supersymmetric\footnote{Alternatively, we can interpret $S^4$ in some neighbourhood of $S^2$ as a bundle $S^2\times I$,over $S^2$, where $I$ is some interval, and encode the restricted dimensions in a larger gauge group $I\otimes G$ living on that bundle. In this framework, a 4d vector multiplets split into 2d chiral and vector multiplets, and the ``unusual coupling'' is encoded in the representation of the chiral multiplet under the new gauge group. The restricted fields are obtained by considering the section of the bundle corresponding to $S^2$}.  As shown in appendix \ref{restriction}, this is the most general renormalizable coupling between the 2d and 4d fields compatible with the symmetries of the theory.

\subsection{Field content and action}

We now proceed to describe the action for theory. It separates into a two dimensional part and a four dimensional one, in the generic form
\begin{align}
S=S_{S^4}+S_{S^2}=\int_{S^4} d^4x\sqrt{g_{S^4}}\mc L_{S^4}+\int_{S^2} d^2x\sqrt{g_{S^2}}\mc L_{S^2}.
\end{align}
The four-sphere has a radius $r$, and we describe it using the stereographic coordinates $x^\mu$, $\mu=1,\cdots,4$, centered at the north pole. The metric is conformally flat, $g_{\mu\nu}=e^{2\Omega}\delta_{\mu\nu}$, with conformal factor $e^{-\Omega}=1+\frac{x^\mu x^\mu}{r^{2}}$.
The two-sphere is taken to be the subspace $x^3=x^4=0$, and we use the coordinate system $x^i$, $i=1,\cdots,2$, induced by inclusion. Our conventions for spinors are described in appendix \ref{Conventions}.

We begin with the description of the vector multiplet on $S^4$\footnote{Our conventions on $S^4$ mostly follow \cite{Hama:2012bg}, and those on $S^2$ mostly follow \cite{Doroud:2012xw}}. The multiplet consists of a gauge field $A_\mu$, a $SU(2)_R$ R-symmetry doublets of Weyl fermions $\lambda_A$, $\bar \lambda_A$, a pair of scalars $\phi$, $\bar \phi$, and a R-symmetry triplet of auxiliary fields $D_{AB}=D_{BA}$. All fields are in the adjoint representation of the gauge group $G$, with Lie algebra $\mg g$. The $SU(2)_R$ indices $\{A,B=1,2\}$ are  raised and lowered by the antisymmetric matrices $C$ and $\tilde C$, defined by $C_{21}=\tilde C^{12}=1$, in the form $\lambda_A=C_{AB}\lambda^B$, $\lambda^A=\tilde C^{AB}\lambda_B$. The Lagrangian for the multiplet is \cite{Hama:2012bg}
\begin{align}
\mc L_{S^4}&=\tfrac1{g^2}\text{Tr}\Big(
\tfrac12F_{\mu\nu}F^{\mu\nu}-4 \mc D_\mu\bar\phi \mc D^\mu\phi-\tfrac8{r^2}\bar\phi\phi-2i\lambda^A\sigma^\mu \mc D_\mu\bar\lambda_A-\tfrac12D^{AB}D_{AB}\nn
&\qquad
-2\lambda^A[\bar\phi,\lambda_A]+2\bar\lambda^A[\phi,\bar\lambda_A]
+4[\bar\phi,\phi]^2
\Big).
\end{align}
Here $F_{\mu\nu}$ is the field strength for $A_\mu$, and $\mc D_\mu=\nabla_\mu-iA_\mu$ is the gauge covariant derivative.

On the two dimensional side, the defect consists in a set of chiral multiplets, or equivalently a single chiral multiplet in a representation $R=\bigoplus_IR_I$ of the gauge group $G$ , decomposing into a direct sum of irreducible representations (flavors) $R_I$. A chiral multiplet in a representation $R$ consists of a scalar $\chi$, a Dirac fermion $\psi$, and an auxiliary scalar $F$ (the corresponding fields are $\bar\chi$, $\bar\psi$, $\bar F$ for the antichiral multiplet). The multiplet is characterized by a mass $m$ and a R-charge $q$, where $m$, $q$ are matrix valued in the flavor space and take constant values $m_I$, $q_I$ in each irreducible representation. As discussed before, the coupling to the $S^4$ vector multiplet takes the same form as the coupling to a $S^2$ vector multiplet. Therefore we can write the Lagrangian as \cite{Doroud:2012xw}
\begin{align}
\mc L_{S^2}&=\bar \chi\l(
-\mc D_i^2+\sigma_1^2+\sigma_2^2+iD+M^2+(2M-\tfrac ir)\sigma_2
\r)\chi
-i\bar\psi\l(\gamma^i \mc D_i-\sigma_1-i\sigma_2\gamma^3-iM\gamma^3\r)\psi\nn
&\qquad+\bar FF+i\bar\psi\lambda\chi-i\bar\chi\bar\lambda\psi,
\end{align}
where $M=m+\tfrac i{2r}q$ and $\mc D_i=\nabla_i-iA_i$. The fields $\sigma_1$, $\sigma_2$, $D$, $\lambda$, $\bar\lambda$, and $A_i$ and form a vector multiplet representation of $\mc N=(2,2)$ supersymmetry. Their exact expression is obtained by decomposing the vector multiplet in terms of representations of the $\mc N=(2,2)$ supersymmetry algebra, which is done by comparing the supersymmetry transformations of the multiplets (see appendix \ref{restriction}). The identification of the vector fields is trivial, and the scalars are given by
\begin{align}
\sigma_1=i(\phi+\bar\phi),&\quad 
\sigma_2=\phi-\bar\phi,\nn
D=-D_{12}-\tfrac 1r&(\phi-\bar\phi)-iF_{34}.
\end{align}

\subsection{Supersymmetry transformations}

Supersymmetry transformations in curved space are most naturally obtained as a subset of the superconformal transformations. Such transformations are parametrized by a set of conformal Killing spinors. Each of these spinors can be expressed as a linear combination of a basis of conformal Killing spinors, each associated to one of the supercharges. In this paper, we describe supersymmetry and superconformal symmetry through conformal Killing spinors, i.e. through the realization of the algebra on the fields.

For generic conformal Killing spinors, the commutator of superconformal transformations contains conformal transformations through conformal Killing vectors. Supersymmetry is obtained by restricting to a maximal subset of the conformal Killing spinors generating only Killing vectors, i.e. isometries. 

On $S^4$, the $\mc N=2$ superconformal transformations are given in terms of $SU(2)_R$ doublets of conformal Killing spinors $\epsilon^A$, $\bar \epsilon^A$. The fields of the vector multiplet transform as
\begin{align}\label{vectorSUSY4d}
(\delta_\epsilon+\delta_{\bar\epsilon}) A_\mu&=i\epsilon^A\sigma_{g\mu}\bar\lambda_A-i\bar\epsilon^A\bar\sigma_{g\mu}\lambda_A,\nn
(\delta_\epsilon+\delta_{\bar\epsilon})\phi&=-i\epsilon^A\lambda_A,\nn
(\delta_\epsilon+\delta_{\bar\epsilon})\bar\phi&=i\bar\epsilon^A\bar\lambda_A,\nn
(\delta_\epsilon+\delta_{\bar\epsilon})\lambda_A&=
\frac12\sigma_g^{\mu\nu}\epsilon_AF_{\mu\nu}+2\sigma_g^\mu\bar\epsilon_A \mc D_\mu\phi+\sigma_g^\mu \nabla_\mu\bar\epsilon_A\phi+2i\epsilon_A[\phi,\bar\phi]+D_{AB}\epsilon^B,\nn 
(\delta_\epsilon+\delta_{\bar\epsilon}) \bar\lambda_A&=
\frac12\bar\sigma_g^{\mu\nu}\bar\epsilon_AF_{\mu\nu}+2\bar\sigma_g^\mu\epsilon_A\mc D_\mu\bar\phi
+\bar\sigma_g^\mu \nabla_\mu\epsilon_A\bar\phi-2i\bar\epsilon_A[\phi,\bar\phi]+D_{AB}\bar\epsilon^B,\nn
(\delta_\epsilon+\delta_{\bar\epsilon}) D_{AB}&=-2i\bar\epsilon_{(B}\bar\sigma_g^\mu\mc D_\mu\lambda_{A)}
+2i\epsilon_{(A}\sigma_g^\mu\mc D_\mu\bar\lambda_{B)}-4[\phi,\bar\epsilon_{(A}\bar\lambda_{B)}]+4[\bar\phi,\epsilon_{(A}\lambda_{B)}].
\end{align}
For $\mc N=2$ supersymmetry, the allowed set of spinors is given in terms of constant spinors $\epsilon_0^A$, $\bar\epsilon_0^A$, by
\begin{align}
\epsilon^A=e^{\tfrac12\Omega}\l(\epsilon_0^A+\tfrac 1{2r}x^\mu\sigma^\mu(\tau^3)^A_{~B}\bar\epsilon_0^B\r),\qquad
\bar\epsilon^A=e^{\tfrac12\Omega}\l(\bar\epsilon_0^A-\tfrac 1{2r}x^\mu\bar\sigma^\mu(\tau^3)^A_{~B}\epsilon_0^B\r).
\end{align}
(See appendix \ref{s4susy}.) The four constant spinors correspond to the eight supercharges of the theory. In the presence of a $S^2$ defect, half of the supersymmetries are broken, and the unbroken ones correspond to spinors of definite ``two dimensional chirality''. The restriction takes the form:
\begin{align}
(-i\sigma^{12})\epsilon_0^1=+\epsilon_0^1,\qquad
(i\bar\sigma^{12})\bar\epsilon_0^1=-\bar\epsilon_0^1,\nn
(-i\sigma^{12})\epsilon_0^2=-\epsilon_0^2,\qquad
(i\bar\sigma^{12})\bar\epsilon_0^2=+\bar\epsilon_0^2.
\end{align}

On $S^2$, the $\mc N=(2,2)$ superconformal transformations are generated by a pair of conformal Killing spinors $\varepsilon$, $\bar \varepsilon$. The fields of the chiral (and antichiral) multiplet transform as
\begin{align}
(\delta_\varepsilon+\delta_{\bar\varepsilon})\chi&=\bar\varepsilon\psi,\nn
(\delta_\varepsilon+\delta_{\bar\varepsilon})\bar\chi&=\varepsilon\bar\psi,\nn
(\delta_\varepsilon+\delta_{\bar\varepsilon})\psi&=i\l(
\gamma_g^i\mc D_i\chi+(\sigma_1-i(\sigma_2+m)\gamma^3)\chi+\tfrac q2\chi\gamma_g^i\nabla_i\r)\varepsilon+F\bar\varepsilon\nn
(\delta_\varepsilon+\delta_{\bar\varepsilon})\bar\psi&=i\l(
\gamma^i\mc D_i\bar\chi+(\sigma_1+i(\sigma_2+m)\gamma^3)\bar\chi+\tfrac q2\bar\chi\gamma_g^i\nabla_i\r)\bar\varepsilon+\bar F\varepsilon\nn
(\delta_\varepsilon+\delta_{\bar\varepsilon})F&=-i\l(\mc D_i\psi\gamma_g^i+\sigma_1\psi-i(\sigma_2+m)\psi\gamma^3+\lambda\chi+\tfrac q2\psi\gamma_g^i\nabla_i
\r)\varepsilon,\nn
(\delta_\varepsilon+\delta_{\bar\varepsilon})\bar F&=-i\l(\mc D_i\bar\psi\gamma_g^i+\bar\psi\sigma_1+i\bar\psi(\sigma_2+m)\gamma^3+\bar\chi\bar\lambda+\tfrac q2\bar\psi\gamma_g^i\nabla_i
\r)\bar\varepsilon.
\end{align}
For supersymmetry, the set of allowed spinors is given in terms of a pair of constant spinors $\varepsilon_0$, $\bar\varepsilon_0$, by
\begin{align}
\varepsilon=e^{\tfrac12\Omega}\l(\varepsilon_0+\tfrac1{2r}x^i\gamma^i\gamma^3\varepsilon_0\r),\qquad
\bar\varepsilon=e^{\tfrac12\Omega}\l(\bar\varepsilon_0-\tfrac1{2r}x^i\gamma^i\gamma^3\bar\varepsilon_0\r).
\end{align}
(See appendix \ref{s2susy}.)
For the theory being considered, the supercharges on $S^2$ and $S^4$ are the same, so the allowed conformal Killing spinors are related. In components, the relation is
\begin{align}
(\epsilon^1)_1=\tfrac 1{\sqrt 2}\varepsilon_1,\quad
(\epsilon^2)_2=\tfrac 1{\sqrt 2}\bar\varepsilon_2,\quad
(\bar\epsilon^1)^2=-\tfrac i{\sqrt 2}\varepsilon_2,\quad
(\bar\epsilon^2)^1=\tfrac i{\sqrt 2}\bar\varepsilon_1.
\end{align}
(See appendix \ref{s2s4susy}.)

\section{Localization}
\label{sec:loc}

The goal of this paper is to compute the partition function 
\begin{align}
Z=\int \mc D\Phi_g \mc D\Phi e^{-S_{2d}[\Phi]-S_{4d}[\Phi]-S_{g}[\Phi_g,\Phi]},
\end{align}
where $\Phi$ denotes the set of fields of the chiral and vector multiplets, $\Phi_g$ is a set of ghosts, $S=S_{2d}+S_{4d}$ is the action defined in the previous section, and $S_g$ is a gauge fixing action. We perform the path integral over a contour in which the bosons satisfy the reality conditions
\begin{align}
 A_\mu^\dagger=A_\mu,\quad
 \phi^\dagger=-\bar\phi,\quad
D_{AB}^\dagger=-D^{AB},\quad
\chi^\dagger=\bar\chi,\quad
F^\dagger=\bar F.
\end{align}

We compute the partition function using supersymmetric localization \cite{Pestun:2007rz,Witten:1988ze}. The method relies on the fact that given a supercharge $\mc Q$, only the $\mc Q$-invariant field configurations contribute to the path integral. Indeed, if the orbit of $\mc Q$ is non-trivial, then the path integral over that orbit vanishes. In practice, we can simplify the computation by deforming the action by a $\mc Q$-exact term in the form $S\to S+t\mc Q\cdot V$. Under certain assumptions, the path integral is not affected by such deformation. If the deformation term is non-negative, we can take the limit $t\to\infty$. In that limit, the saddle-point approximation becomes exact, and thus can be used to compute the partition function for the original theory exactly. Schematically, the path integral reduces to a sum (or integral) over the set $F$ of zeros of $\mc Q\cdot V$, and a Gaussian integral, in the form
\begin{align}
Z=\sum_{\Phi_0\in F}e^{-S[\Phi_0]}\int \mc D\Phi e^{-t(\mc Q\cdot V)[\Phi-\Phi_0]_{\text{Quad}}}
=\sum_{\Phi_0\in F}e^{-S[\Phi_0]}Z_{1-\text{loop}}[\Phi_0].
\end{align}
The one-loop partition function $Z_{1-\text{loop}}[\Phi_0]$ is a ratio of functional determinants, and is greatly simplified by supersymmetry. Indeed, as mentioned before, only the supersymmetric configurations contribute, the others canceling pairwise. We compute it using the Atiyah-Singer index theorem. Among the set $F$ of classical configurations, we distinguish the contribution of instantons from the set $F_0$ of configurations of zero instanton number. In the $t\to\infty$ limit, there is no smooth classical field configuration with instantons, but there are still singular instanton configurations localized at the poles \cite{Pestun:2007rz}. These instanton configurations do not affect the one-loop determinant, and their contribution at each pole is given by an instanton partition function. We can thus expect the partition function to take the form
\begin{align}
Z=\sum_{\Phi_0\in F_0}e^{-S[\Phi_0]}Z_{1-\text{loop}}[\Phi_0]\l|Z_{\text{inst}}[\Phi_0]\r|^2.
\end{align}

\subsection{The supercharge and deformation terms}

To perform the localization, we choose a supercharge compatible with the localization for both the two and four dimensional fields, i.e. a supercharge of the unbroken supersymmetry. We pick the supercharge $\mc Q$ such that the poles are left invariant by $\mc Q^2$. In that case $\mc Q^2$ generates a $U(1)$ group, which consists of a combination of $SO(2)\subset SO(3)$ and $SO(2)_\perp$ rotations. In terms of conformal Killing spinors in the two dimensional formalism, the constraint is $\varepsilon_0\gamma^i\bar\chi_0+\bar\varepsilon_0\gamma^i\chi_{0}=0$ (i.e. the coefficient of $M_{i5}$ must vanish in the Killing vector (\ref{2dKilling})), and is satisfied by imposing $\gamma^3\varepsilon_{0}=+\varepsilon_{0}$, $\gamma^3\bar\varepsilon_{0}=-\bar\varepsilon_{0}$. This leaves two independent spinors, and we choose the combination defined by $\varepsilon_0=i\gamma^1\bar\varepsilon_0$. In components, we write $(\varepsilon_0)_1=i(\bar\varepsilon_0)_2=\varepsilon_{\mc Q}$, where $\varepsilon_{\mc Q}$ is the parameter for the transformation $\delta_{\mc Q}=r^{-\frac12}\varepsilon_{\mc Q}\mc Q$. The square of the supercharge $\mc Q$ is realized as
\begin{align}
\mc Q^2&=M_{12}+M_{34}+\tfrac12R+\mc G[\Lambda]-irm,\nn
\Lambda&=-iv\cdot A+r(f(x)\sigma_1-i\sigma_2),\nn
 f(x)&=e^{\Omega}(1-\tfrac{x^\mu x^\mu}{4r^{2}})=\cos(\theta),
\end{align}
where $v$ is the Killing vector associated with $M_{12}M_{34}$, $\theta$ measures the angle on $S^4$ relative to the north pole, and $m$ is the mass of the multiplet.

This choice of supercharge is compatible with previous computations on $S^2$ and $S^4$, hence we can pick a similar deformation term, and the computations follow a similar pattern. In particular, the classical and one-loop computations are almost the same, the main difference being in the classical field configuration appearing in the two-dimensional one-loop determinant. We review these computations and adapt them to the present case, mostly following the conventions of \cite{Doroud:2012xw} on $S^2$, and those of \cite{Hama:2012bg} on $S^4$. However, the instanton partition function for the four dimensional theory is affected by the presence of the defect, and must be computed separately.

We deform the action by a non-negative $\mc Q$-exact term $\mc Q V=\mc Q V_{4d}+\mc Q V_{2d}$. The two-dimensional part $\mc Q V_{2d}$ can be taken to be the action $S_{S^2}$ itself, as it is $\mc Q$-exact. On the four-dimensional side, we take
\begin{align}
V=\text{Tr}[(\mc Q\lambda_A)^\dagger\lambda_A+(\mc Q\bar\lambda_A)^\dagger\bar\lambda_A].
\end{align}
The resulting deformation term $\mc Q V_{4d}$ manifestly satisfies the required properties.

\subsection{Classical configurations and one-loop determinants}

For the vector multiplet, the classical configurations are given by the zeros of the deformation term $\mc Q V_{4d}$, for which the bosonic part is
\begin{align}
\mc Q V_{4d}^{\text{bos}}=\text{Tr}[(\mc Q\lambda_A)^\dagger(\mc Q\lambda_A)+(\mc Q\bar\lambda_A)^\dagger(\mc Q\bar\lambda_A)].
\end{align}
The classical configurations thus coincide with the supersymmetric configurations. Up to a gauge transformation, the smooth solutions to $\mc Q V_{4d}^{\text{bos}}=0$ are given in terms of a constant $\mg g$-valued parameter $a$, by \cite{Pestun:2007rz}
\begin{align}
A_\mu=0,\quad
\phi=-\bar\phi=\tfrac 1{2r}a,\quad
D_{12}=-\tfrac 1{r^2}a,\quad
D_{11}=D_{22}=0.
\end{align}
For the chiral multiplet, the equation of motion for $D_{12}$ and the above configuration fix $\chi=0$, and $F$ must vanish by its own equation of motion. The action for such configuration is $S_{cl}(a)=\tfrac{8\pi^2}{g^2}\text{Tr}(a_0^2)$.

The one-loop determinants for the chiral and vector multiplets are computed independently from each other, although they both depend on the background of the vector multiplet. For a generic vector multiplet background, the one-loop determinant for the chiral multiplet is a product over the weights of $R$:
\begin{align}
\tilde Z_{1-\text{loop}}^{2d}(\Lambda)=\prod_{w\in R}\frac{\Gamma(\omega\cdot\Lambda_N-irM)}{\Gamma(1-\omega\cdot\Lambda_S+irM)}
\end{align}
where $N$, $S$ denote the values at the north and south poles. This result is computed in appendix \ref{one-loop}. For the above background, the formula reduces to
\begin{align}\label{2d1loop}
Z_{1-\text{loop}}^{2d}(a)=\prod_{w\in R}\frac{\Gamma(-i\omega\cdot a-irM)}{\Gamma(1+i\omega\cdot a+irM)}
\end{align}
The one-loop determinant for the vector multiplet is (see \cite{Pestun:2007rz})
\begin{align}
\tilde Z_{1-\text{loop}}^{4d}(a)=\prod_{\alpha\in\Delta}G(1+ia\cdot\alpha)G(1-ia\cdot\alpha),
\end{align}
where $\Delta$ is the set of roots of $\mg g$, and $G(z)$ is the Barnes $G$-function \cite{Barnes01011901}. 

\subsection{The instanton partition function}

In \cite{Pestun:2007rz}, it was shown that the instanton contribution to the partition function at each pole in the absence of defect is given by Nekrasov's instanton partition function \cite{Nekrasov:2003rj} of the theory in the $\Omega$-background on $\mb R^4$. The argument given in that paper is still valid here, however the presence of the defect modifies the instanton partition function, which is now that of a similar theory with a $\mb R^2$ defect, i.e. the $\mb R^4$ version of the theory considered in this paper in the $\Omega$-background. We leave the computation of the exact instanton partition function $Z^{2d-4d}_{\text{inst}}(a)$ for future work.

Summing up the previous computations, we write the complete partition function as
\begin{align}
Z_{2d-4d}=\int_{\mg h}dae^{-\tfrac{8\pi^2}{g^2}\text{Tr}(a^2)}Z^{2d}_{1-\text{loop}}(a)Z^{4d}_{1-\text{loop}}(a)|Z^{2d-4d}_{\text{inst}}(a)|^2,
\end{align}
where we reduced the integral over $a$ to the Cartan subalgebra $\mg h$ at the price of a Jacobian factor $\prod_{\alpha\in\Delta}(\alpha\cdot a)$, inserted in the modified one-loop determinant
\begin{align}\label{4d1loop}
Z_{1-\text{loop}}^{4d}(a)=\prod_{\alpha\in\Delta}(\alpha\cdot a)G(1+ia\cdot\alpha)G(1-ia\cdot\alpha),
\end{align}

\section{Generalizations}
\label{sec:general}

In this section we consider the generalization of the above results, where we allow additional types of multiplets. Namely, we consider the addition of hypermultiplets on $S^4$ and vector multiplets on the defect. As in the previous case (see appendix \ref{restriction}) symmetry and renormalizability imposes heavy constraints on the possible couplings between the defect and the bulk. A vector multiplet on the defect cannot be coupled consistently with a 4d field, and a hypermultiplet can only couple to the defect through a (heavily constrained) superpotential. However, a superpotential term does not affect the partition function, so the localization procedure involves no more couplings between the bulk and the defect than the one considered previously.

A vector multiplet on the defect cannot be coupled consistently with a 4d field, but a hypermultiplet can couple to the defect in two different ways. The first one is through a superpotential: the restricted hypermultiplet appears as two series of chiral multiplets, which can appear in the superpotential. By renormalizability the superpotential must be at most linear in the 4d fields, and only the top chiral multiplet in each series is allowed (i.e. without transverse derivative), and gauge invariance constrains the superpotential further. In any case, a superpotential term does not affect the partition function, so it is irrelevant for the present computation. The other possible coupling is through a four dimensional vector multiplet frozen to its vacuum expectation value \cite{Gomis:2014eya}: if a Lagrangian is invariant under some flavor symmetry, we can weakly gauge it by introducing a introducing a vector multiplet, then freezing it.


The localization procedure for the generalized theory is unchanged from the one described previously. Here we consider vector multiplets $V_{4d}$ and  $V_{2d}$, with gauge groups $G_{4d}$ and  $G_{2d}$. $V_{2d}$ is also associated to a Fayet-Iliopoulos parameter $\xi$ and a topological angle $\theta$, appearing in the combination $\tau=\tfrac{\theta}{2\pi}+i\xi$. The matter on $S^4$ forms a hypermultiplet $H$ with mass $m_H$, in a representation $R_H$ of $G_{4d}$\footnote{As in the 2d case, hypermultiplet masses are obtained through a four dimensional vector multiplet frozen to its vacuum expectation value. In principle, one could also couple such vector multiplet to chiral fields on the defect, but the effect is equivalent to giving (equal) twisted masses to the 2d fields \cite{Gomis:2014eya}. Namely, one can obtain such coupling from others simply by constraining masses, so we do not need to consider it here.}. The chiral multiplet $\Psi$ on the defect has mass and R-charge $M_\Psi=m_\Psi+\tfrac i{2r}q_{\Psi}$, and is in a representation $R_{\Psi}$ of $G_{4d}\times G_{2d}$. Localization works exactly as before, and we obtain the formula 
\begin{align}\label{finalResult}
Z_{2d-4d}=\sum_B\int_{\mg h_{2d}}d\tilde a e^{-4\pi i r \text{Im}\text{Tr}[\tau (\tilde a+\tfrac i{2r}B)]}
\int_{\mg h_{4d}}dae^{-\tfrac{8\pi^2}{g^2}\text{Tr}(a^2)}
Z_{1-\text{loop}}(a,\tilde a, B)|Z^{2d-4d}_{\text{inst}}(a,\tilde a, B)|^2,
\end{align}
\begin{align}
\text{where}\quad Z_{1-\text{loop}}(a,\tilde a, B)&=Z^{V_{2d}}_{1-\text{loop}}(\tilde a, B)Z^{V_{4d}}_{1-\text{loop}}(a)
Z^{H}_{1-\text{loop}}(a)Z^{\Psi}_{1-\text{loop}}(a,\tilde a, B),
\end{align}
\begin{align}
\text{and}\quad Z^{\Psi}_{1-\text{loop}}(a,\tilde a, B)&=\prod_{w\in R}\frac{\Gamma(-i\omega\cdot (a,\tilde a+\tfrac i{2r}B)-irM)}{\Gamma(1+i\omega\cdot (a,\tilde a+\tfrac i{2r}B)+irM)}.
\end{align}
The other one-loop determinants are unchanged from the separate expressions on $S^2$ and $S^4$  \cite{Doroud:2012xw,Benini:2012ui,Pestun:2007rz}, and as before we leave the computation of the instanton partition function for future work. 

\acknowledgments{This work was supported by the Perimeter Institute for Theoretical Physics and the Natural Sciences and Engineering Research Council of Canada (NSERC).  Research at Perimeter Institute is supported by the Government of Canada through Industry Canada and by the Province of Ontario through the Ministry of Economic Development and Innovation. The author is thankful to N. Doroud, D. Gaiotto, and J. Gomis for useful discussions and remarks.

\begin{appendix}

\section{Coordinates and spinors}
\label{Conventions}

We use the stereographic coordinates on $S^4$. The coordinates are labelled $x^\mu$, $\mu=1,\cdots,4$, and the metric is conformally flat:
\begin{align}
g_{\mu\nu}=e^{2\Omega}\delta_{\mu\nu},\qquad e^{-\Omega}=1+\frac{x^\mu x^\mu}{4r^{2}},
\end{align}
where $r$ is the radius of the sphere. By $x^\mu x^\mu$, it is understood that the contraction is performed using the flat space metric, $x^\mu x^\mu=\delta_{\mu\nu}x^\mu x^\nu$. The $S^2$ subspace is taken to be along the 1-2 plane. When needed, we split the coordinates and indices into parallel and orthogonal pairs, and write the indices as $i,j=1,2$, and $\tilde i,\tilde j=3,4$. The coordinates $x^i$ on $S^2$ are given by inclusion. The induced metric is $g_{ij}=e^{2\Omega}\delta_{ij}$, with conformal factor $e^{-\Omega}=1+\tfrac{x^i x^i}{4r^{2}}$.

On $S^4$, we use the Weyl spinor formalism. The matrices $(\sigma^\mu_g)_{\alpha\dot\alpha}$, $(\bar\sigma^{\mu}_g)^{\dot\alpha\alpha}$ satisfy $\{\sigma^{\mu}_g,\bar\sigma^{\nu}_g\}=2g^{\mu\nu}$. Since the space is conformally flat, they are simply related to the flat space matrices $\sigma^\mu$, $\bar\sigma^{\mu}$ (anticommuting to $\delta_{\mu\nu}$), by $\sigma^{\mu}_g=e^{-\Omega}\sigma^{\mu}$, $\bar\sigma^{\mu}_g=e^{-\Omega}\bar\sigma^{\mu}$. We take a basis in which the flat space matrices are given by $\sigma^4=\bar\sigma^4=1$ and $\sigma^m=-\bar\sigma^m=-i\tau^m$, $m=1,2,3$, where $\tau^m$ are the Pauli matrices. Spinor indices are raised and lowered by the charge conjugation matrices $C$ and $\tilde C$ in the form $\lambda_\alpha=C_{\alpha\beta}\lambda^\beta$, $\lambda^\alpha=\tilde C^{\alpha\beta}\lambda_\beta$ (and similarly for right-handed indices). The charge conjugation matrices are antisymmetric and satisfy $C_{\alpha\gamma}\tilde C^{\gamma\beta}=\delta_\alpha^\beta$. By convention we take $C_{21}=\tilde C^{12}=-C_{12}=-\tilde C^{21}=1$.

The two dimensional spinors are taken to be Dirac spinors, and the Dirac matrices $(\gamma_g^{i})_a^{~b}$ satisfy the Clifford algebra $\{\gamma_g^{i},\gamma_g^{j}\}=2g^{ij}$. As in the four dimensional case, they can be expressed in terms of the flat space Dirac matrices $\gamma^{i}$ as $\gamma_g^{i}=e^{-\Omega}\gamma^{i}$. The chirality matrix is $\gamma^{3}=-i\gamma^{1}\gamma^{2}$. We take a basis in which $\gamma^{m}$ $(m=1,2,3)$ are numerically equal to the Pauli matrices $\tau^{m}$. Spinor indices are raised and lowered as four dimensional spinors.

\section{Conformal Killing spinors and supersymmetry}
\label{Spinors}

\subsection{Conformal Killing spinors on $S^4$ and $S^2$}

Conformal Killing spinor $\epsilon$, $\bar\epsilon$ in four dimensions are solutions of the equations
\begin{align}
 \nabla_\mu\epsilon=\sigma_{g\mu}\bar\epsilon',\qquad
\nabla_\mu\bar\epsilon'=-\tfrac{1}{4r^2}\bar\sigma_{g\mu}\epsilon,\nn
 \nabla_\mu\bar\epsilon=\bar\sigma_{g\mu}\epsilon',\qquad
\nabla_\mu\epsilon'=-\tfrac{1}{4r^2}\sigma_{g\mu}\bar\epsilon,
\end{align}
where $\epsilon'$, $\bar\epsilon'$ are some auxiliary spinors.
The solutions to these equations are
\begin{align}
\epsilon=e^{\tfrac12\Omega}(\epsilon_0+x^\mu\sigma^\mu\bar\epsilon_1),\qquad
\bar\epsilon=e^{\tfrac12\Omega}(\bar\epsilon_0+x^\mu\bar\sigma^\mu\epsilon_1),
\end{align}
where $\epsilon_0$, $\epsilon_1$, $\bar\epsilon_0$, $\bar\epsilon_1$ are arbitrary constant spinors, and can be obtained using the flat space $(r\to\infty)$ solution together with Weyl covariance. 

Conformal Killing spinors in two dimensions work exactly as in four dimension. In this case, the equations are
\begin{align}
 \nabla_i\varepsilon=\gamma_{g i}\varepsilon',\qquad
\nabla_i\varepsilon'=-\tfrac{1}{4r^2}\gamma_{g i}\varepsilon,
\end{align}
and are solved by
\begin{align}
\varepsilon=e^{\tfrac12\Omega}(\varepsilon_0+x^i\gamma^i\varepsilon_1)
\end{align}

\subsection{$\mc N=2$ supersymmetry on $S^4$}
\label{s4susy}

The supersymmetry algebra on $S^4$ is most conveniently obtained as a subalgebra of the superconformal algebra, in which the spacetime transformations are restricted to the isometries. In the following we describe this process through the realization of the algebra on fields and spacetime.

$\mc N=2$ superconformal transformations are realized through $SU(2)$ doublets of conformal Killing spinors $\epsilon^A$, $\bar\epsilon^A$. For the vector multiplet, the superconformal transformations are given by (\ref{vectorSUSY4d}), and the algebra is realized on the vector multiplet as 
\begin{align}
[\delta_{\bar\epsilon}+\delta_{\bar\epsilon},\delta_\eta+\delta_{\bar\eta}]&=\mc L_v +\mc G(-i v\cdot A+\Phi)+\omega\Omega+\tilde\Theta \tilde R+\Theta_{AB}R^{AB},
\end{align}
where $\mc L_v$ is a lie derivative,$\mc G$ is a gauge transformation, $\Omega$ is a Weyl transformation, and $\tilde R$, $R^{AB}$ are $U(1)$ and $SU(2)$ $R$-symmetry transformations. The various parameters are given by
\begin{align}\label{superconformalS4}
v^\mu&=2i\epsilon^A\sigma_g^\mu\bar\eta_A-(\epsilon\leftrightarrow\eta),\qquad 
\omega=\tfrac14\nabla_\mu v^\mu,\nn
\tilde\Theta&=\tfrac i4(\epsilon^A\sigma_g^\mu\nabla_\mu\bar\eta_A-\nabla_\mu \epsilon^A\sigma_g^\mu\bar\eta_A)-(\epsilon\leftrightarrow\eta),\nn
\Theta_{AB}&= i(\epsilon_{(A}\sigma_g^\mu\nabla_\mu\bar\eta_{B)}-\nabla_\mu \epsilon_{(A}\sigma_g^\mu\bar\eta_{B)})-(\epsilon\leftrightarrow\eta),\nn
\Phi&=-4\epsilon^A\eta_A\bar\phi+4\bar\epsilon^A\bar\eta_A\phi
\end{align}
In particular, the spacetime transformations are generated by the conformal Killing vector $v$. To restrict to supersymmetry, we restrict the set of allowed conformal Killing spinors in such a way that $v$ is a Killing vector, i.e. it generates only isometries of the sphere. We expand

\begin{align}
v&=2i\epsilon_0^A\sigma^\mu\bar\eta_{0A}\partial_\mu
+2i(\epsilon_0^A\eta_{1A}-\bar\epsilon_1^A\bar\eta_{0A})x^\mu\partial_\mu
-2i(\epsilon_0^A\sigma^{\mu\nu}\eta_{1A}+\bar\epsilon_1^A\bar\sigma^{\mu\nu}\bar\eta_{0A})x_\mu\partial_\nu\nn
&\qquad+2i\bar\epsilon_1^A\bar\sigma^\mu\eta_{1A}(x^2\partial_\mu-2x_\mu x^\nu\partial_\nu)-(\epsilon\leftrightarrow\eta)
\nn
&=-2\epsilon_0^A\sigma^\mu\bar\eta_{0A}P_\mu
+2(\epsilon_0^A\eta_{1A}-\bar\epsilon_1^A\bar\eta_{0A})D
+(\epsilon_0^A\sigma^{\mu\nu}\eta_{1A}+\bar\epsilon_1^A\bar\sigma^{\mu\nu}\bar\eta_{0A})M_{\mu\nu}\nn
&\qquad-2\bar\epsilon_1^A\bar\sigma^\mu\eta_{1A}K_\mu-(\epsilon\leftrightarrow\eta),
\end{align}
where $P_\mu$, $D$, $M_{\mu\nu}$, $K_\mu$ are the generators of the conformal transformations in the scalar representation. The $SO(5,1)$ symmetry can be made manifest by defining
\begin{align}\label{SO6}
M_{\mu5}=nP_\mu-mK_\mu,\qquad
M_{\mu6}=nP_\mu+mK_\mu,\qquad
M_{56}=D,
\end{align}
where $n=\cosh\alpha$, $m=\sinh\alpha$ for some hyperbolic angle $\alpha$.
\begin{align}
v&=
-\l[\tfrac1n\epsilon_0^A\sigma^\mu\bar\eta_{0A}+\tfrac1m\bar\epsilon_1^A\bar\sigma^\mu\eta_{1A}\r]M_{\mu5}
-\l[\tfrac1n\epsilon_0^A\sigma^\mu\bar\eta_{0A}-\tfrac1m\bar\epsilon_1^A\bar\sigma^\mu\eta_{1A}\r]M_{\mu6}
\nn&\qquad
+(\epsilon_0^A\sigma^{\mu\nu}\eta_{1A}+\bar\epsilon_1^A\bar\sigma^{\mu\nu}\bar\eta_{0A})M_{\mu\nu}
+2(\epsilon_0^A\eta_{1A}-\bar\epsilon_1^A\bar\eta_{0A})M_{56}-(\epsilon\leftrightarrow\eta)
\end{align}
For the sphere, the isometry group is $SO(5)$, which is made manifest as a subgroup of $SO(5,1)$ by setting $\tanh\alpha=\tfrac mn=\tfrac1{4r^2}$. For this choice of angle, the generators $M_{\mu6}$ and $M_{56}$ are not allowed. This imposes the conditions
\begin{align}
(\epsilon_1^A\sigma^\mu\bar\eta_{1A}-\eta_1^A\sigma^\mu\bar\epsilon_{1A})
&=-\tfrac1{4r^2}(\epsilon_0^A\sigma^\mu\bar\eta_{0A}-\eta_0^A\sigma^\mu\bar\epsilon_{0A}),\nn
\epsilon_0^A\eta_{1A}+\epsilon_1^A\eta_{0A}&=\bar\epsilon_1^A\bar\eta_{0A}+\bar\epsilon_0^A\bar\eta_{1A},
\end{align}
which are satisfied by
\begin{align}
\epsilon_1^A=\tfrac1{2r}(\tau^3)^A_{~B}\epsilon_0^B,&\qquad
\bar\epsilon_1^A=-\tfrac1{2r}(\tau^3)^A_{~B}\bar\epsilon_0^B,\nn
\text{or}\qquad
\nabla_\mu\epsilon^A=-\tfrac1{2r}\sigma_{g\mu}(\tau^3)^A_{~B}\epsilon^B&,\qquad
\nabla_i\bar\epsilon^A=\tfrac1{2r}\sigma_{g\mu}(\tau^3)^A_{~B}\bar\epsilon^B.
\end{align}
(and similarly for $\eta^A$, $\bar\eta^A$). After imposing these conditions, we obtain $\mc N=2$ supersymmetry, and the Killing vector is given by
\begin{align}
v&=
-\tfrac2n(\epsilon_0^A\sigma^\mu\bar\eta_{0A}-\eta_0^A\sigma^\mu\bar\epsilon_{0A})M_{\mu5}
-\tfrac1r(\tau^3)^B_{~A}(\epsilon_0^A\sigma^{\mu\nu}\eta_{0B}+\bar\epsilon_0^A\bar\sigma^{\mu\nu}\bar\eta_{0B})M_{\mu\nu}.
\end{align}
The other parameters in (\ref{superconformalS4}) reduce to 
\begin{align}
\omega&=\tilde\Theta=\Theta_{11}=\Theta_{22}=0,\nn
\Theta_{12}&=2\Theta=-\tfrac{2i}r(\epsilon^A\eta_A-\bar\epsilon^A\bar\eta_A),\nn
\Phi&=-4\epsilon^A\eta_A\bar\phi+4\bar\epsilon^A\bar\eta_A\phi,
\end{align}
and the algebra simplifies to
\begin{align}
[\delta_{\bar\epsilon}+\delta_{\bar\epsilon},\delta_\eta+\delta_{\bar\eta}]&=\mc L_v +\mc G(-i v\cdot A+\Phi)+\Theta R,
\end{align}
where $R=4R^{12}$ is the unbroken $R$-charge.

The coupling of the theory to matter on a $S^2$ surface breaks the spacetime symmetry of the theory to $SO(3)\times SO(2)_\perp$. Here $SO(3)$ represents the isometries of $S^2$, i.e. rotations and translations on the 1-2 plane, and $SO(2)_\perp$ represents the rotations on the 3-4 plane. Hence to obtain the set of unbroken supersymmetries, we need a subset of the Killing spinors which does not generate $M_{\tilde i5}$ or $M_{i\tilde j}$. This requirement imposes the conditions
\begin{align}
\epsilon_0^A\sigma^{\tilde i}\bar\eta_{0A}-\eta_0^A\sigma^{\tilde i}\bar\epsilon_{0A}=0,\qquad
\epsilon_0^A\sigma^{i\tilde j}\eta_{0B}+\bar\epsilon_0^A\bar\sigma^{i\tilde j}\bar\eta_{0B}=0,
\end{align}
which can be satisfied by requiring the spinors to have a definite "two dimensional chirality". We choose them to satisfy
\begin{align}
(-i\sigma^{12})\epsilon_0^1=+\epsilon_0^1,\qquad
(-i\bar\sigma^{12})\bar\epsilon_0^1=-\bar\epsilon_0^1,\nn
(-i\sigma^{12})\epsilon_0^2=-\epsilon_0^2,\qquad
(-i\bar\sigma^{12})\bar\epsilon_0^2=+\bar\epsilon_0^2
\end{align}
(and similarly for $\eta^A$, $\bar\eta^A$). The Killing spinor then simplifies to
\begin{align}
v
&=\tfrac{4}n(\epsilon_0^{[1}\sigma^i\bar\eta_{0}^{2]}-\eta_0^{[1}\sigma^i\bar\epsilon_{0}^{2]})M_{i5}
+\tfrac{4i}r\l(
\epsilon_0^{[1}\eta_0^{2]}-\bar\epsilon_0^{[1}\bar\eta_0^{2]}
\r)M_{12}+\tfrac{4i}r\l(
\epsilon_0^{[1}\eta_0^{2]}+\bar\epsilon_0^{[1}\bar\eta_0^{2]}
\r)M_{34}
\end{align}

\subsection{$\mc N=(2,2)$ supersymmetry on $S^2$}
\label{s2susy}

For supersymmetry on $S^2$, we proceed exactly as in the $S^4$ case. $\mc N=(2,2)$ supersymmetry are realized through a pair of real conformal Killing spinors $\varepsilon$, $\bar\varepsilon$, and the algebra is realized as
\begin{align}
[\delta_{\bar\varepsilon}+\delta_{\bar\varepsilon},\delta_\chi+\delta_{\bar\chi}]&=\mc L_v +\mc G(-i v\cdot A+\Phi)+\omega\Omega+\Theta R+\tilde\Theta \mc A+\alpha m,
\end{align}
where $R$ and $\mc A$ are $R$-symmetry transformations, $m$ is the mass of the multiplet, and the parameters are
\begin{align}
v^\mu&=-i\varepsilon\gamma_g^i\bar\eta-(\varepsilon\leftrightarrow\chi),\qquad 
\omega=\tfrac14\nabla_\mu v^\mu,\nn
\Theta&=-\tfrac i4(\nabla_i\varepsilon\gamma_g^i\bar\chi-\varepsilon\gamma_g^i\nabla_i\bar\chi)-(\varepsilon\leftrightarrow\chi)\nn
\tilde\Theta&= \tfrac i4(\nabla_i\varepsilon\gamma^3\gamma_g^i\bar\chi-\varepsilon\gamma^3\gamma_g^i\nabla_i\bar\chi)-(\varepsilon\leftrightarrow\chi),\nn
\Phi&=i\varepsilon\bar\chi\sigma_1-\varepsilon\gamma_g^3\bar\chi\sigma_2-(\varepsilon\leftrightarrow\chi),\quad
\alpha=-\varepsilon\gamma^3\bar\chi-(\varepsilon\leftrightarrow\chi).
\end{align}
The conformal Killing vector $v$ expands as
\begin{align}
v&=\varepsilon_0\gamma^i\bar\chi_{0}P_i
-(\varepsilon_0\bar\chi_{1}-\varepsilon_1\bar\chi_{0})D
-\tfrac12(\varepsilon_0\gamma^{ij}\bar\chi_{1}+\varepsilon_1\gamma^{ij}\bar\chi_{0})M_{ij}
+\varepsilon_1\gamma^i\bar\chi_{1}K_i-(\varepsilon\leftrightarrow\chi)\nn
&=
\tfrac 12\l[\tfrac1n\varepsilon_0\gamma^i\bar\chi_0-\tfrac1m\varepsilon_1\gamma^i\bar\chi_{1}\r]M_{i5}
+\tfrac 12\l[\tfrac1n\varepsilon_0\gamma^i\bar\chi_0+\tfrac1m\varepsilon_1\gamma^i\bar\chi_{1}\r]M_{i6}\nn
&\qquad-\tfrac12(\varepsilon_0\gamma^{ij}\bar\chi_{1}+\varepsilon_1\gamma^{ij}\bar\chi_{0})M_{ij}
-(\varepsilon_0\bar\chi_{1}-\varepsilon_1\bar\chi_{0})M_{56}-(\varepsilon\leftrightarrow\chi),
\end{align}
where we used the restriction of (\ref{SO6}) to make the $SO(3,1)$ symmetry manifest. We reduce the symmetry group to $SO(3)$ by imposing 
\begin{align}\label{SUSY2d}
\varepsilon_1=\tfrac 1{2r}\gamma^3\varepsilon_0&,\qquad
\bar\varepsilon_1=-\tfrac 1{2r}\gamma^3\bar\varepsilon_0,\nn
\text{or}\qquad
\nabla_i\varepsilon=+\tfrac1{2r}\gamma_{gi}\gamma^3\varepsilon&,\qquad
\nabla_i\bar\varepsilon=-\tfrac1{2r}\gamma_{gi}\gamma^3\bar\varepsilon.
\end{align}
Under these constraints, the Killing vector is
\begin{align}\label{2dKilling}
v&=\tfrac1n(\varepsilon_0\gamma^i\bar\chi_0+\bar\varepsilon_0\gamma^i\chi_{0})M_{i5}
+\tfrac 1{2r}(\varepsilon_0\gamma^{ij}\gamma^3\bar\chi_{0}-\bar\varepsilon_0\gamma^{ij}\gamma^3\chi_{0})M_{ij}\nn
&=\tfrac1n(\varepsilon_0\gamma^i\bar\chi_0+\bar\varepsilon_0\gamma^i\chi_{0})M_{i5}
+\tfrac i{r}(\varepsilon_0\bar\chi_{0}-\bar\varepsilon_0\chi_{0})M_{12},
\end{align}
and the other parameters of the algebra simplify to 
\begin{align}
\omega=\tilde\Theta=0,\quad \Theta=\tfrac i{2r}\alpha
\end{align}

\subsection{Relating the spinor formalisms}
\label{s2s4susy}

We now proceed to relate the two spinor formalisms introduced for $\mc N=(2,2)$ supersymmetry. This amounts to matching the conformal Killing spinors on both sides. We recall that on one side we have four four dimensional Weyl spinors $\epsilon_0^A$, $\bar\epsilon_0^A$ of definite two dimensional chirality, and on the other we have two unconstrained two dimensional Dirac spinors $\varepsilon_0$, $\bar\varepsilon_0$. To proceed, we write the spinors in components in chiral bases, which gives four elementary spinors on both sides:
\begin{align}
(\epsilon_0^1)_1,(\epsilon_0^2)_2, (\bar\epsilon_0^1)^2,(\bar\epsilon_0^2)^1,\quad\text{and}\quad
(\varepsilon_0)_1,(\varepsilon_0)_2, (\bar\varepsilon_0)_1,(\bar\varepsilon_0)_2.
\end{align}
The matching must respect the supersymmetry algebra, i.e. the Killing vector $v^\mu$ must be preserved on $\mc S^2$. This can be achieved by setting
\begin{align}
(\epsilon_0^1)_1=\tfrac 1{\sqrt 2}(\varepsilon_0)_1,\quad
(\epsilon_0^2)_2=\tfrac 1{\sqrt 2}(\bar\varepsilon_0)_2,\quad
(\bar\epsilon_0^1)^2=-\tfrac i{\sqrt 2}(\varepsilon_0)_2,\quad
(\bar\epsilon_0^2)^1=\tfrac i{\sqrt 2}(\bar\varepsilon_0)_1.
\end{align}
This form allows to relate the nonzero components of the conformal Killing spinors by 
\begin{align}
(\epsilon^1)_1=\tfrac 1{\sqrt 2}\varepsilon_1,\quad
(\epsilon^2)_2=\tfrac 1{\sqrt 2}\bar\varepsilon_2,\quad
(\bar\epsilon^1)^2=-\tfrac i{\sqrt 2}\varepsilon_2,\quad
(\bar\epsilon^2)^1=\tfrac i{\sqrt 2}\bar\varepsilon_1.
\end{align}

\section{Decomposition of the vector multiplet on $S^4$}
\label{restriction}

In this section we decompose the $\mc N=2$ vector multiplet into representations of the $\mc N=(2,2)$ superconformal algebra. We focus on the $\mc N=(2,2)$ vector multiplet, for which an exact expression is needed to the coupling to the defect. 
Our goal is to find a set of fields $(A_i, \sigma_1,\sigma_2,D,\lambda,\bar\lambda)$ among these which transform as a $\mc N=(2,2)$ vector multiplet, in the form
\begin{align}
\delta_\varepsilon A_i&=-\tfrac i2(\bar\varepsilon\gamma_{gi}\lambda+\varepsilon\gamma_{gi}\bar\lambda),\nn
\delta_\varepsilon \sigma_1&=\tfrac 12(\bar\varepsilon\lambda-\varepsilon\bar\lambda),\nn
\delta_\varepsilon \sigma_2&=-\tfrac i2(\bar\varepsilon\gamma^{3}\lambda+\varepsilon\gamma^{3}\bar\lambda),\nn
\delta_\varepsilon \lambda&=(\tfrac12\gamma_g^{ij}F_{ij}-\gamma^3\gamma_g^i\mc D_i\sigma_2+i\gamma_g^i\mc D_i\sigma_1
-\gamma^3[\sigma_1,\sigma_2]-D)\varepsilon+i(\sigma_1+i\sigma_2\gamma^3)\gamma_g^i\nabla_i\varepsilon,\nn
\delta_\varepsilon \bar\lambda&=(\tfrac12\gamma_g^{ij}F_{ij}-\gamma^3\gamma_g^i\mc D_i\sigma_2-i\gamma_g^i\mc D_i\sigma_1
-\gamma^3[\sigma_1,\sigma_2]+D)\bar\varepsilon-i(\sigma_1-i\sigma_2\gamma^3)\gamma_g^i\nabla_i\varepsilon,\nn
\delta_\varepsilon D&=
\tfrac i2\mc D_i(\varepsilon\gamma_g^i\bar\lambda-\bar\varepsilon\gamma_g^i\lambda)
-\tfrac i2\l[\sigma_1,\varepsilon\bar\lambda+\bar\varepsilon\lambda
\r]
+\tfrac12\l[\sigma_2,\varepsilon\gamma^3\bar\lambda-\bar\varepsilon\gamma^3\lambda
\r]
.
\end{align}
For the gauge field, the identification $A_i^{\mc N=(2,2)}=A_i^{\mc N=2}$ is obvious. We evaluate
\begin{align}
\delta_\epsilon A_i&=i\epsilon^A\sigma_\mu\bar\lambda_A-i\bar\epsilon^A\bar\sigma_\mu\lambda_A\nn
&=\tfrac i{\sqrt 2}\varepsilon^1(\gamma_{gi})_1^{~2}(\lambda_1)_2
+\tfrac 1{\sqrt 2}\varepsilon^2(\gamma_{gi})_2^{~1}(\bar\lambda_1)^1
+\tfrac 1{\sqrt 2}\bar\varepsilon^1(\gamma_{gi})_1^{~2}(\bar\lambda_2)^2
-\tfrac i{\sqrt 2}\bar\varepsilon^2(\gamma_{gi})_2^{~1}(\lambda_2)_1,
\end{align}
which implies the identifications
\begin{align}
(\lambda_1)_2=-\tfrac 1{\sqrt2}\bar\lambda_2,\quad
(\lambda_2)_1=\tfrac1{\sqrt2}\lambda_1,\quad
(\bar\lambda_1)^1=-\tfrac i{\sqrt2}\bar\lambda_1,\quad
(\bar\lambda_2)^2=-\tfrac i{\sqrt2}\lambda_2.
\end{align}
Under these identifications, the scalars $\phi$, $\bar\phi$ transform as
\begin{align}
\delta_\epsilon \phi&=-i\epsilon^A\lambda_A=\tfrac i2\varepsilon^2\bar\lambda_2-\tfrac i2\bar\varepsilon^1\lambda_1,\nn
\delta_\epsilon \bar\phi&=i\bar\epsilon^A\bar\lambda_A=\tfrac i2\varepsilon^1\bar\lambda_1-\tfrac i2\bar\varepsilon^2\lambda_2,
\end{align}
implying $\sigma_1=i(\phi+\bar\phi)$, $\sigma_2=\phi-\bar\phi$.
The auxiliary field $D_{12}$ transforms as
\begin{align}
\delta_\epsilon D_{12}&=
-\tfrac i2(\varepsilon\gamma_g^i\mc D_i\bar\lambda-\bar\varepsilon\gamma_g^i\mc D_i\lambda)
+\tfrac i2\l[\sigma_1,\varepsilon\bar\lambda+\bar\varepsilon\lambda
\r]
-\tfrac12\l[\sigma_2,\varepsilon\gamma^3\bar\lambda-\bar\varepsilon\gamma^3\lambda
\r]+i\delta(F_{34})\nn
&=-\delta(D+\tfrac1r\sigma_2
+iF_{34}),
\end{align}
implying the identification
\begin{align}
D=-D_{12}-\tfrac 1r(\phi-\bar\phi)-iF_{34}.
\end{align}
A similar computation can be done to show that the supersymmetry transformations of the fermions are consistent with the above identifications.

\section{Restriction in the superfield formalism and supersymmetric couplings}
\label{restriction}

In this section we present an alternative description of the theory, based on the superfield formalism in flat space. This description has the advantage of making explicit the possible supersymmetric couplings between the 2d and 4d multiplets. It is equivalent to the $S^2\subset S^4$ case by classical conformal invariance, which allows to relate the Lagrangians and supersymmetry transformations in the two cases. 

We first write down the vector multiplet on $\mb R^4$ in terms of representations of $\mc N=(2,2)$ supersymmetry. It takes the form of a chiral multiplet $\Phi=\Phi(x^i,x^{\tilde i},\theta, \bar\theta)$ coupled to a vector multiplet $V=V(x^i,x^{\tilde i},\theta, \bar\theta)$. The symmetry in the directions orthogonal to the defect is encoded in the gauge transformations $\delta_\Lambda V=i(\Lambda-\Lambda^\dagger)$, $\delta_\Lambda \Phi=2\sqrt2\partial_+\Lambda$, where $\Lambda=\Lambda(x^i,x^{\tilde i},\theta, \bar\theta)$ is a chiral field, and $\partial_\pm=\tfrac12(\partial_3\pm i\partial_4)$. The four dimensional Lagrangian
\begin{align}
\mc L_{\mb R^4}=\int d^4\theta\l((\Phi^\dagger-2\sqrt2i\partial_-V)(\Phi+2\sqrt2i\partial_+V)-2\partial_+V\partial_-V\r)
+\int d^4\theta W^2+\int d^4\bar \theta \bar W^2
\end{align}
is gauge invariant, and can be shown to reproduce that of a $\mc N=2$ vector multiplet on $\mb R^4$. The restriction to $\mb R^2$ contains the fundamental fields $\Phi$ and $V$ at $x^{\tilde i}=0$, but also their derivatives on the defect. Therefore the set of restricted fields consists of an infinite tower of vector multiplets $V_R^{(m,n)}=(\partial_+^m\partial_-^nV)(x^{\tilde i}=0)$, and a tower of chiral multiplets $\Phi_R^{(m,n)}=(\partial_+^m\partial_-^n\Phi)(x^{\tilde i}=0)$. In particular, the defect sees infinitely many copies of the gauge group $G$, which we label by $G^{(m,n)}$. Also in the Wess-Zumino gauge the different chiral multiplets mix under supersymmetry.

Given a chiral multiplet $\Psi$, the most general supersymmetric (and gauge invariant) coupling to the vector multiplet is obtained by picking a representation $R^{(m,n)}$ for each of the gauge groups. The Lagrangian may also include a superpotential $\mc W=\mc W(\Psi)$ (The superpotential cannot depend on $\Phi_R^{(m,n)}$ by gauge invariance), and twisted masses. Twisted masses are obtained by coupling the chiral multiplet to a  (2d) vector multiplet frozen to its vacuum expectation value $\tilde V$. In this paper we are interested in a local, renormalizable interaction term, which preserves the $SO(2)_\perp$ symmetry between the transverse directions. Locality requires the interaction to involve only a finite number of multiplets, and by $SO(2)_\perp$ symmetry we must restrict to those with $m=n$. By renormalizability all the representations other than $R^{(0,0)}$ must be trivial. Therefore the most general allowed two dimensional Lagrangian is of the form
\begin{align}\label{R2Lagrangian}
\mc L_{\mb R^2}&=\int d^4\theta~ \Psi^\dagger \exp(V^{(0,0)}_R+\tilde V)\Psi+\int d^2\theta~ \mc W(\Psi)+c.c.
\end{align}

\section{One-loop determinants of the chiral multiplet from index theorem}
\label{one-loop}

We compute the one-loop determinant of the chiral multiplet using the equivariant index theorem for transversally elliptic operators . For our purpose, the statement of the index theorem is as follow \cite{Pestun:2007rz,Atiyah:1974}. Let $E_0$, $E_1$ be a pair of vector bundles over a manifold $M$, and $G$ a compact Lie group acting on the bundles and the manifold. Let $D: \Gamma(E_0)\to\Gamma(E_1)$ map sections of the bundles and commute with the action of $G$. We also require $D$ to be transversally elliptic\footnote{A differential operator $D$ is said to be transversally elliptic if its symbol is invertible for sections of the cotangent bundle $T^*M$ transversal to the $G$-orbit. In a local coordinate frame, the symbol is obtained from the highest order part of $D$ by replacing partial derivatives at each point $x$ by momenta, $\partial_i\to ip_i$, where $\{p_i\}$ is set of coordinates for a point $p$ on $T_x^*M$. For transversal ellipticity, we require the symbol to be invertible for all $p$ in the subspace of $T_x^*M$ transversal to the $G$-orbit at each point $x\in M$ (in any coordinate frame).}. The index of $D$ is defined as
\begin{align}
\text{ind}D(\hat t)=\text{tr}_{\text{Ker}D}\hat t-\text{tr}_{\text{Coker}D}\hat t,
\end{align}
where $\hat t$ is an element of the maximal torus of $G$. For a transversally elliptic operator, the Kernel and Cokernel are both infinite dimensional, but can be decomposed as a sum of irreducible representations of $G$, each appearing with a finite multiplicity. Therefore we can expand the index in formal series. For $G=U(1)$, $\hat t=e^{i\epsilon \hat g}=t^g$, where $\epsilon\in \mb R$, and $g$ is the generator of the Lie algebra $u(1)$, and we can expand $\text{ind}D(\hat t)=\sum_ic_it^{w_i}$. However, the expansion is not unique, and some care must be taken in choosing the appropriate expansion. Assuming that $G$ has a discrete set of fixed points $F$, the index theorem gives the index of $D$ as a sum over $F$:
\begin{align}
\text{ind}D(\hat t)=\sum_{p\in F}\frac{\text{tr}_{E_0(p)}\hat t-\text{tr}_{E_1(p)}\hat t}{\det_{T_pM}(1-\hat t)}.
\end{align}
To use the index theorem, we first build a $\mc Q$-complex from the fields. We write the fields of each multiplet as a pair of (sets of) bosons $\{\Phi,\tilde\Phi\}$ and a pair of (sets of) fermions $\{\Psi,\tilde\Psi\}$, such that $\tilde\Psi=\mc Q\Phi$, $\tilde\Phi=\mc Q\Psi$. The fields are sections of vector bundles, which we write $\{E_\Phi,E_{\tilde\Phi}\}$, $\{E_\Psi,E_{\tilde\Psi}\}$. Then $\mc Q^2$ maps each bundle to itself, and we can take $G$ to be the $U(1)$ Lie group generated by $ \mc Q^2$. The quadratic part of the deformation term $\mc V_q$ in the form
\begin{align}
\mc V_q=\l(\begin{array}{ll}
\tilde\Psi&\Psi
\end{array}\r)\cdot
\l(\begin{array}{ll}
D_{00}&D_{01}\\
D_{10}&D_{11}
\end{array}\r)\cdot
\l(\begin{array}{l}
\Phi\\\tilde\Phi
\end{array}\r)=\bf\Psi^\dagger\cdot\bf D\bf \cdot\Phi.
\end{align}
This construction gives a smooth linear map $D_{10}:\Gamma(E_{\Phi})\to\Gamma(E_{\Psi})$. In the following, we assume that $D_{10}$ commutes with the action of $\mc Q$, allowing us to use the index theorem. The quadratic part of the Lagrangian can then be written as 
\begin{align}
\mc Q\mc V_q=(\mc Q\bf\Psi)\cdot\bf D\bf \cdot\Phi+\bf\Psi\cdot\bf D\bf \cdot(\mc Q\Phi)
=\bf\Phi\cdot\boldsymbol{\mc Q}_R^2\cdot\bf D\bf \cdot\Phi+\bf\Psi\cdot\bf D\bf \cdot\boldsymbol{\mc Q}^2\cdot\Psi,
\end{align}
where $\boldsymbol{\mc Q}^2=\text{diag}(1,\mc Q^2)$, $\boldsymbol{\mc Q}_R^2=\text{diag}(\mc Q_R^2,1)$, and $\mc Q_R$ is the supercharge $\mc Q$ acting on the right. Then for real fields, the one-loop determinant can be written as 
\begin{align}
Z_{1-\text{loop}}=\sqrt{\frac{\det(\bf D\bf \cdot\boldsymbol{\mc Q}^2)}{\det(\boldsymbol{\mc Q}_R^2\cdot\bf D)}}
=\sqrt{\frac{\det_{\bf \Psi}\mc Q^2}{\det_{\bf\Phi}\mc Q^2}}
=\sqrt{\frac{\det_{\text{Coker} D_{10}}\mc Q^2}{\det_{\text{Ker} D_{10}}\mc Q^2}},
\end{align}
where the last equality follows from the assumption that $D_{10}$ commutes with $\mc Q$, therefore it relates the action of $\mc Q$ on both bundles outside its kernel and cokernel. For complex fields, square root is absent. The ratio of determinant can be obtained from the equivariant index through
\begin{align}
\text{ind}D_{10}(t)=\sum_ic_it^{w_i}\Longleftrightarrow
\frac{\det_{\text{Coker} D_{10}}\mc Q^2}{\det_{\text{Ker} D_{10}}\mc Q^2}=\prod_i w_i^{-c_i}.
\end{align}

\subsection{The chiral multiplet}

We now proceed to compute the one-loop determinant of the chiral multiplet using the above method, as done in \cite{Benini:2012ui}. Here we assume that the operator $D_{10}$ is transversally elliptic, and avoid computing its exact form. We take the $\mc Q$-complex defined by $\Phi=\phi$, $\varepsilon_{\mc Q}\Psi=\varepsilon\gamma^3\psi$. The two fixed points are the north and south poles. Near the north pole $\mc Q^2$ acts in complex coordinates as $\mc Q^2(z,\bar z)=(z,-\bar z)$, so the denominator is $(1-t)(1-t^{-1})$, and similarly at the south pole. The numerator is obtained from the action of $\mc Q^2$ at the poles, obtained from
\begin{align}
\mc Q^2\phi_N&=[\Lambda_N-irM]\phi_N,&
\mc Q^2\phi_S&=[\Lambda_S-irM]\phi_S,\nn
\mc Q^2\varepsilon\psi_N&=[\Lambda_N-irM-1]\varepsilon\psi_N,&
\mc Q^2\varepsilon\psi_S&=[\Lambda_S-irM-1]\varepsilon\psi_S,
\end{align}
where $\Lambda=-iv\cdot A+r(f(x)\sigma_1-i\sigma_2)$, $M=m+\tfrac i{2r}q$. Here the fields of the vector multiplet are treated as background fields. More formally, the bundles $E_\Phi$, $E_\Psi$ are isomorphic to $K_{q,m}\otimes\mc R$ and $S\otimes K_{q-1,m}\otimes\mc R$, where $K_{q,m}$ is a one-dimensional vector bundle encoding a R-charge $q$ and a mass $m$, $\mc R$ is a vector bundle transforming in the representation $R$ of $G$, and $S$ is a (one-dimensional) subbundle of a Spin-$\tfrac12$ bundle interpolating between definite angular momenta $-\tfrac12$ at the north pole and $+\tfrac12$ at the south pole. The index evaluates to a sum over the roots of $R$:
\begin{align}
\text{ind}D_{10}(t)=\l[\sum_{w\in R}
\frac{t^{\omega\cdot\Lambda_N-irM}}{1-t}
\r]_N
+\l[\sum_{w\in R}
\frac{t^{\omega\cdot\Lambda_S-irM}}{1-t}
\r]_S
\end{align}
The series expansion at each pole is dictated by the symbol of $D_{10}$ (see \cite{Pestun:2007rz} for details). Here we need to expand the north pole contribution in powers of $t$, and the south pole contribution in powers of $t^{-1}$, giving
\begin{align}
\text{ind}D_{10}(t)=\sum_{w\in R}\sum_{n=0}^{\infty}\l(t^{n+\omega\cdot\Lambda_N-irM}-(t^{-1})^{n+1-\omega\cdot\Lambda_S+irM}\r)
\end{align}
From this index we deduce the one-loop determinant (up to an irrelevant phase)
\begin{align}
Z_{1-\text{loop}}=\prod_{w\in R}\frac{\prod_{n=0}^{\infty}(1+n-\omega\cdot\Lambda_S+irM)}{\prod_{n=0}^{\infty}(n+\omega\cdot\Lambda_N-irM)}.
\end{align}
We regularize the products according to the formula
\begin{align}
\prod_{n=0}^{\infty}(n+m)\to\frac1{\Gamma(m)},
\end{align}
giving the result
\begin{align}
Z_{1-\text{loop}}=\prod_{w\in R}\frac{\Gamma(\omega\cdot\Lambda_N-irM)}{\Gamma(1-\omega\cdot\Lambda_S+irM)}.
\end{align}

\end{appendix}

\bibliographystyle{JHEP}
\bibliography{irr}

\end{document}